\documentclass[aps,prb,floatfix,reprint,superscriptaddress,showpacs]{revtex4-1}
\usepackage{epsfig}
\usepackage{amsmath}
\usepackage{amsfonts}
\usepackage{amssymb}
\usepackage{mathptmx}
\usepackage{eucal}

\DeclareMathOperator{\Tr}{\mathop{\mathrm{Tr}}}
\DeclareMathOperator{\sgn}{\mathop{\mathrm{sgn}}}
\DeclareMathOperator{\re}{\mathop{\mathrm{Re}}}
\DeclareMathOperator{\im}{\mathop{\mathrm{Im}}}
\newcommand{\oM}{ \overline{M} }
\newcommand{\oN}{ \overline{N} }
\newcommand{\tM}{ \widetilde{M} }
\newcommand{\Eq}[1]{Eq.~(\ref{#1})}
\newcommand{\Eqs}[1]{Eqs.~(\ref{#1})}

\begin{document}

\title{Dissipative charge transport in diffusive superconducting
double-barrier junctions}

\author{E.~V.~Bezuglyi}
\email{eugene.bezuglyi@gmail.com}
\author{E.~N.~Bratus'}
\affiliation{B.Verkin Institute for Low Temperature Physics and Engineering,
61103 Kharkov, Ukraine}
\author{V.~S.~Shumeiko}
\affiliation{Chalmers University of Technology, S-41296 G\"oteborg, Sweden}


\begin{abstract}
We solve the coherent multiple Andreev reflection (MAR) problem and calculate
current-voltage characteristics (IVCs) for Josephson SINIS junctions, where S
are local-equilibrium superconducting reservoirs, I denotes tunnel barriers,
and N is a short diffusive normal wire, the length of which is much smaller
than the coherence length, and the resistance is much smaller than the
resistance of the tunnel barriers. The charge transport regime in such
junctions qualitatively depends on a characteristic value $\gamma = \tau_d
\Delta$ of relative phase shifts between the electrons and retro-reflected
holes accumulated during the dwell time $\tau_d$. In the limit of small
electron-hole dephasing $\gamma \ll 1$, our solution recovers a known formula
for a short mesoscopic connector extended to the MAR regime.  At large
dephasing, the subharmonic gap structure in the IVC scales with
$\gamma^{-1}$, which thus plays the role of an effective tunneling parameter.
In this limit, the even gap subharmonics are resonantly enhanced, and the IVC
exhibits portions with negative differential resistance.
\end{abstract}

\maketitle

\section{Introduction \label{Intro}}

Accurate theoretical description of nonequilibrium charge transport in
Josephson junctions is an important and active research field. The concept of
Multiple Andreev Reflections (MAR) \cite{KBT} is a universal framework
explaining the nature of dissipative current in different types of junctions.
The quasiparticles injected in the junction at applied voltage smaller than
the superconducting energy gap, $eV<2\Delta$, can only escape into reservoirs
at zero temperature by multiple traversing the junction due to repeated
Andreev reflections, each time gaining energy $eV$. Such a process generates
coherent transfer of multiple electron charge $ne= 2\Delta/V$ across the
junction. Such a mechanism is important at small temperatures when the single
particle tunneling is exponentially weak, and the Andreev transport becomes
dominant. The most complete quantitative MAR theory has been developed for
ballistic contacts. The central element here is the solution for a single
conducting channel with given transmissivity, which enters the sum over
conducting channels in the net current. Various approaches  for constructing
such a solution have been developed based on the scattering theory, tunneling
Hamiltonian, and quasiclassical Green's
functions.\cite{Arnold,Zaikin,MAR95-1,MAR95-2,Cuevas96}

Extension of the theory has been suggested for disordered weak links, e.g.,
diffusive constrictions and super\-con\-ductor-insulator-superconductor (SIS)
tunnel junctions with disordered insulating layers. In this case the
summation over the channels is performed by taking into account the
distribution of random transmission eigenvalues for corresponding structure
in the normal state.\cite{Bardas,Naveh,BG} This method, however, is only
valid for short supercon\-duc\-ting weak links, in which the dwell time,
i.e., the time of quasiparticle diffusion along the whole MAR staircase, is
small on a quantum time scale defined by the inverse or\-der parameter
$\Delta^{-1}$. In junctions with large dwell time, e.g., long
super\-con\-ductor-normal metal-super\-con\-ductor (SNS) junctions and
double-barrier SINIS junctions containing Breit-Wig\-ner resonances, the
method does not work because of strong energy dispersion of the scattering
amplitudes. The electron- and hole-like quasiparticles acquire different
scattering phases during propagation through the junction (electron-hole
dephasing), which results in the single-chan\-nel MAR current being dependent
on the scattering amplitudes rather than the scattering
coefficients.\cite{Ingermann2001,Cuevas97,Johansson99} Thus the distribution
of transmission eigenvalues becomes irrelevant and is to be replaced by the
statistics of scattering amplitudes, which is generally unknown except of
some particular cases.\cite{Samuelsson2002}

In order to investigate the MAR problem in diffusive junctions with large
electron-hole dephasing, one has to directly solve quasiclassical
Keldysh-Green's function equations.\cite{LO} This task, however, is
technically demanding because a nonlinearity of equations and nonstationary
character of the MAR problem lead to hardly tractable two-dimensional
infinite chains of Green's function harmonics. So far the problem has been
analytically solved for superconducting con\-stric\-tions,\cite{Zaitsev} and
for the opposite case of very long SNS and SINIS junctions, where the
Josephson current is completely suppressed and a incoherent MAR theory can be
formulated\cite{Bezugly2000,Bezuglyi2001} analogous to the one for ballistic
junctions.\cite{OTBK} A perturbative scheme for a superconducting film
interrupted by a tunnel junction has been suggested,\cite{Bezuglyi2006}
however no general methods to analytically treat the problem in SNS junctions
of intermediate lengths exists up to now, and even numerical solution
presents a difficult and time consuming task.\cite{Cuevas06}

In this paper we present a study of the junction for which the coherent MAR
problem can be relatively easily solved and fully investigated both
analytically and numerically for the entire range from weak to strong
electron-hole dephasing. We consider a SINIS junction with opaque NIS
interfaces having equal resistances $R$ much larger than the resistance $R_N$
of the diffusive normal wire, the length $2d$ of which is supposed to be much
smaller than the coherence length $\sqrt{\mathcal{D}/\Delta}$ ($\mathcal{D}$
is the diffusion coefficient, $\hbar =  1$). In such junctions, the dwell
time $\tau_d$ is characterized by the
parameter\cite{Bezugly2000,Volkov,Brinkman,3T}
\begin{eqnarray} \label{gamma}
\gamma = \tau_d\Delta = \frac{R}{R_N}\frac{\Delta}{E_{\text{Th}}},
\end{eqnarray}
where $E_{\text{Th}} = \mathcal{D}/(2d)^2$ is the Thouless energy. This
parameter defines the amplitude of the Josephson current and the magnitude of
the minigap in the energy spectrum within the normal wire $\Delta_g
\sim{\Delta }/(1+\gamma)$. In short junctions, the Thouless energy is large
compared to $\Delta$, therefore, the parameter $\gamma$ may have arbitrary
value depending on the junction length and transparency. Correspondingly, the
value of the minigap may vary between $0$ and $\Delta$ reflecting the
crossover from large to small electron-hole dephasing.

We construct the MAR solution of the Kel\-dysh-Green's function equations and
evaluate the dc current for arbitrary $\gamma$ by means of the second-order
recurrences similar to that for ballistic structures.\cite{MAR95-1,MAR95-2}
For the case of small electron-hole dephasing ($\gamma\ll 1$) we show that
the current is given by a general formula for mesoscopic connector
\cite{Nazarov99} [see \Eq{Nazarov} below], i.e., the average of the result
for the single channel \cite{MAR95-1,MAR95-2} over the distribution of
transmission eigenvalues for a double-barrier normal diffusive
structure.\cite{Naz2B} This result coincides with the result of
Ref.~\onlinecite{BG} and it corresponds to a broad Breit-Wigner resonance in
the single channel. In the opposite case of large elec\-t\-ron-hole
dephasing, $\gamma\gg 1$, the current can be explicitly presented as a sum of
multiparticle tunnel currents which scale with the effective tunneling
parameter $\gamma^{-1}$.

The structure of the paper is as follows. We formulate basic equations and
construct analytical solutions in Section \ref{Basic}. In Section
\ref{Numerics}, we present and discuss the results of numerical calculation
of the current-voltage characteristics (IVCs). The multiparticle currents in
the limit $\gamma \gg 1$ are calculated analytically in Section
\ref{Analytics}, which also includes evaluation of the excess current.

\section{Basic equations and solution \label{Basic}}

We start our quantitative consideration with equation for the Keldysh-Green's
function $\check{G}(x,t_1,t_2)$ in the normal wire ($-d < x < d$),
\begin{equation}
\left[\sigma_z \hat{E}, \check{G}\right] = i\mathcal{D}\partial_x
\left(\check{G}
\partial_x \check{G}\right), \quad \check{G}^2 = 1,\quad \check{G} =
\begin{pmatrix} \hat{g}^R & \hat{G}^K \\ 0 & \hat{g}^A
\end{pmatrix}. \label{Keldysh}
\end{equation}
Here, $\hat{g}^{R,A}$ are the retarded/advanced Green's functions, $\hat{f}$
is the matrix distribution function, $\hat{G}^K =\hat{g}^R \hat{f} - \hat{f}
\hat{g}^A$, and the kernel of the energy operator $\hat{E}$ is ${E}(t_1,t_2)
= i\partial_{t_1} \delta(t_1 -t_2)$. All products in \Eq{Keldysh} are time
convolutions: $(A B)(t_1,t_2) = \int dt A(t_1,t)B(t,t_2) $. The electric
current is defined as
\begin{equation} \label{I1}
I(t) = (\pi{g_N}/{4e}) \Tr \tau_K \left(\check{G} \partial_x
\check{G}\right)(t,t) ,\quad \tau_K=\sigma_z \tau_x
\end{equation}
where $g_N$ is the conductance of the normal wire per unit length, and the
$\sigma$ and $\tau$ matrices operate in the Nambu and the Keldysh space,
respectively. At the tunnel barriers $x=\pm d$, we apply the
Ku\-p\-ri\-yanov-Lukichev's boundary conditions\cite{KL}
\begin{align}\label{KL}
g_N\left(\check{G} \partial_x \check{G}\right)_{\pm d} &= \pm
(2R)^{-1}\left[\check{G}_{\pm d}, \check{G}_{R,L}\right].
\end{align}
The equilibrium Keldysh-Green's functions $\check{G}_{R,L}$ in the right and
left superconducting electrodes are constructed with the local-equilibrium
Green's and distribution functions
\begin{subequations} \label{gf}
\begin{align}
&\hat{g}_{R,L}=\sigma_z u(\epsilon_\pm)+i\exp({\pm i\sigma_z eVt})\sigma_y
v(E),\label{g}
\\
&\hat{f}_{R,L}=f(\epsilon_\pm), \quad f(E)=\tanh\frac{E}{2T},\quad
\epsilon_\pm = E\pm \sigma_z \frac{eV}{2},
\\
&\!\!\!\!\!u^{R,A}(E) = \frac{E}{\xi},\;\; v^{R,A}(E) = \frac{\Delta}{\xi},
\;\; \xi^{R,A} =\sqrt{(E\pm i0)^2-\Delta^2},\label{gRL}
\end{align}
\end{subequations}
given in the $(E,t)$-representation: $A(E,t)=\int {d\tau}
e^{iE\tau}A(t+\tau/2,t-\tau/2)$. In \Eqs{gf} we use the antisymmetric gauge
of the superconducting phase $\phi_R = -\phi_L = eVt$, satisfying the
Josephson relation $\phi=\phi_R-\phi_L = 2eVt$.

Solution of \Eqs{Keldysh}--\eqref{gf}, being generally difficult, essentially
simplifies in short junctions with opaque barriers. Averaging \Eq{Keldysh}
along the wire and using \Eqs{KL} and \eqref{gamma}, we get
\begin{equation}
2\gamma \bigl[\sigma_z\hat{E}, \overline{\check{G}}\bigr] =
i\Delta\left(\left[\check{G}_d, \check{G}_R\right] + \left[\check{G}_{-d},
\check{G}_L\right]\right), \label{HG}
\end{equation}
where $\overline{\check{G}}$ denotes spatially averaged value of
${\check{G}}$. In the tunnel limit $R \gg R_N$, the Kel\-dysh-Green's
functions are approximately spatially homogeneous within the normal
wire\cite{Volkov,3T} $\overline{\check{G}} \approx \check{G}_d \approx
\check{G}_{-d}$. Thus, denoting these quantities by a single notation
$\check{G}$, we arrive at the commutator equation
\begin{equation} \label{Comm}
[\check{A}, \check{G}] =0, \quad \check{A}= \check{G}_+ -i\sigma_z\tau_d
\hat{E}, \quad \check{G}_\pm = \frac{1}{2}\left(\check{G}_R \pm \check{G}_L
\right).
\end{equation}
A similar approach has been used in analysis of current transport in a NINIS
structure.\cite{Samuel} Following Refs.~\onlinecite{Samuel} and
\onlinecite{Solution}, we find a physically relevant solution of \Eq{Comm}
satisfying the normalization condition $\check{G}^2 = 1$ in \Eq{Keldysh},
\begin{equation} \label{FormalSol}
\check{G} = \frac{\check{A}}{\sqrt{\check{A}^2}} = \frac{1
}{\pi}\int_{-\infty}^\infty d\lambda\, \check{K}(\lambda), \qquad
\check{K}(\lambda)=(\check{A}+i\lambda)^{-1}.
\end{equation}
Applying \Eq{KL} to \Eq{I1} and using \Eq{FormalSol}, we symmetrize the
quantity $I(t)$ with respect to the left and right reservoirs,
\begin{equation} \label{I2}
I(t) =  \int_{-\infty}^\infty  \frac{d\lambda}{8 eR} \Tr \tau_K
\left[\check{K}(\lambda), \check{G}_- \right](t,t).
\end{equation}

The structure of the matrix current $ \left[\check{K}(\lambda), \check{G}_-
\right]$ in \Eq{I2} is quite similar to the solution of the MAR problem for a
single ballistic channel with the transparency $D=(\lambda^2+1)^{-1}$ given
in Ref.~\onlinecite{Zaitsev} and differs from the latter by an additional
term $-i\sigma_z\tau_d \hat{E}$ in the matrix $\check{K}$, which describes
the electron-hole dephasing during the dwell time $\tau_d$. If this effect is
negligibly small, $\gamma \to 0$, \Eq{I2} rewritten in terms of the functions
$G_{L,R}$ and the transparency variable $D$ can be transformed to the known
formula for a short connector\cite{Nazarov99} generalized to the
nonstationary case of voltage biased $\text{SINIS}$ junction,
\begin{align} \label{Nazarov}
&I(t) = \frac{\pi}{8 eR}\int_0^1 dD\, \Tr \tau_K \frac{D\rho(D)
\bigl[\check{G}_L,\check{G}_R \bigr]} {1 +
\frac{D}{4}\bigl(\bigl\{\check{G}_L,\check{G}_R\bigr\}-2\bigr)}(t,t),
\\
&\rho(D) = \frac{1}{\pi D^{3/2}\sqrt{1-D}}, \qquad \int_0^1D\rho(D)\, dD = 1.
\label{rho}
\end{align}
The fact that the MAR current in this limit is given by a convolution of
non-resonant single channel current with the transparency distribution
$\rho(D)$ for a double-barrier potential\cite{Naz2B} (see also
Refs.~\onlinecite{Naveh} and \onlinecite{Schep}) is consistent with a wide
resonance in the single channel. This result justifies the method and the
result of Ref.~\onlinecite{BG}. We note that in the static limit
$\dot{\phi}=0$, \Eq{FormalSol} reproduces the result\cite{Volkov,3T} of a
direct solution of Usadel equations,
\begin{align}
&\hat{g} = \frac{\sigma_z E +i\sigma_y \Delta(E,\phi)}
{\sqrt{E^2-\Delta^2(E,\phi)}},\label{Solg1}
\end{align}
where $\Delta(E,\phi) = {\Delta\cos(\phi/2)} [1- i\gamma/v(E)]^{-1}$.

In the general case of arbitrary $\gamma$, calculation of the matrix
$\check{K}$ in \Eq{FormalSol} can be performed by expanding all quantities
over the harmonics of $eV$: $A(E,t)=\sum\nolimits_m A(E,m) e^{-imeVt}$. In
this representation, the time averaged (dc) current reads as
\begin{align} \label{I5}
I =\int_{-\infty}^\infty \int_{-\infty}^\infty  \frac{d\lambda\; dE}{16 \pi
eR} \sum_m \Tr \check{K}(\lambda,E,m) \bigl[\check{G}_-(E,-m), \tau_K \bigr],
\end{align}
and the local-equilibrium functions contain only three harmonics, $m=0,\pm
1$. In \Eq{I5} we rearranged the factors in the integrand using the fact that
time averaging is equivalent to the trace in the time domain. After some
algebra, we find the function $\check{G}_+=\sigma_z {G}_0^+ \delta_{m,0}+
i\sigma_y {G}_1^+ \delta_{|m|,1}$ and the commutator
$\left[\check{G}_-,\tau_K \right]=\sigma_z {G}_0^-\delta_{m,0}+ i\sigma_y
{G}_1^- m\delta_{|m|,1}$, where
\begin{align}
&G_0^+=\frac{1}{2} \left[i \left(\oN_+ + \oN_-\right)+N_+F_+ + N_-F_-\right],
\\
&G^-_0=\tau_z(f_+N_+-f_-N_-)+i\tau_y(N_+-N_-),
\\
&G_1^+=\frac{1}{2} \left(i\oM + MF \right),\quad G^-_1=i\oM\tau_x +Mf,
\label{G1+}
\\
&F = \tau_z +2f\tau_+,\quad \tau_+ = (1/2)(\tau_x+i\tau_y), \nonumber
\\
&N=\re u^R,\quad M=\re v^R, \quad \oN=\im u^R,\quad \oM=\im v^R. \nonumber
\end{align}
Here and in the following, the lower indices $\pm$ denote the energy shift by
$\pm eV/2$. The function $N(E)$ is the BCS density of states (normalized over
its value in the normal metal), which turns to zero at $|E|<\Delta$ along
with the function $M(E)$, while the functions $\oN(E)$ and $\oM(E)$ vanish
outside the energy gap. This leads to the following expression for the dc
current,
\begin{align} \label{I6}
&I =   \int_{-\infty}^\infty dE \,J(E) ,\quad J(E)= \int_{-\infty}^\infty
\frac{d\lambda }{16 \pi eR} \,j(E,\lambda),
\\
&j= j_0 + j_1 + j_{-1},
\\
&j_0 = \Tr \check{K}(E,0) \sigma_z{G}_0^-, \quad j_{\pm 1} = \pm\Tr
\check{K}(E,\mp 1) i\sigma_y{G}_1^-. \label{Jm}
\end{align}
Here, $J(E)$ is the current spectral density, whereas the quantity
$j(E,\lambda)$ can be interpreted as a generalized spectral density depending
on the auxiliary parameter $\lambda$.

According to \Eq{FormalSol}, the matrix $\check{K}(\lambda)$ obeys the
equation $(\check{A} +i\lambda)\check{K}(\lambda)=1$; in the
($E,m$)-representation, it has the form
\begin{align}
& \sum\nolimits_{m'} {G_+}[E+(m-m')eV/2, m']{K}(E-eVm'/2,m-m')
\nonumber \\
&+i[\lambda-\sigma_z\tau_d (E+eVm/2)]K(E,m) =\delta_{m,0}\nonumber
\end{align}
(we omit the `check' on top of the $4\times 4$ matrices), where the sum
actually contains only three nonzero terms with $m' = 0,\pm 1$. Introducing
the quantity $K_m(E)=K(E+meV/2,m)$, we obtain the $4\times 4$ matrix
recurrence relation
\begin{align}
&(H_{m} + i\lambda) K_m + h_{m} K_{m-1} + {h}_{m+1} K_{m+1}=\delta_{m,0},
\label{EqK}
\\
&H_{m}= \sigma_z Q_m, \quad Q_m=G_0^+(E_m)-i\tau_d E_m,\nonumber
\\
&h_{m}= i\sigma_y q_m,\quad q_m = G_1^+(E_{m-1/2}), \quad E_m =
E+meV.\nonumber
\end{align}

Solution of \Eq{EqK} can be found by the matrix version of the chain
fractions formalism\cite{MAR95-1,MAR95-2} using the ansatz
\begin{align}\label{SP}
K_m = \begin{cases}  S_m S_{m-1}\ldots S_1 K_0, & m>0, \\  P_m P_{m+1}\ldots
P_{-1} K_0, & m<0. \end{cases}
\end{align}
Recurrence relations for the ``matrix chain fractions'' $S_m$ and $P_m$ with
the boundary conditions $S_m \to 0$ at $m \to \infty$ and $P_m \to 0$ at $m
\to -\infty$ follow from \Eqs{EqK} and \eqref{SP} at $m \neq 0$,
\begin{subequations} \label{recurr1}
\begin{align}
S_m=-(H_{m}+i\lambda+h_{m+1}S_{m+1})^{-1}h_{m},
\\
P_m=-(H_{m}+i\lambda+h_{m}P_{m-1})^{-1}h_{m+1}.
\end{align}
\end{subequations}
At $m=0$ we obtain a nonuniform equation, the solution of which is
$K_0(E)=(H_{0}+i\lambda +h_{0}P_{-1} + h_{1}S_1)^{-1}$. Thus the functions
$K(E,m)$ in \Eq{Jm} read as
\begin{subequations} \label{KK}
\begin{align}
&K(E,0)=K_0(E), \quad K(E,1)=S_1(E_{-})K_0(E_{-}),
\\
&K(E,-1)= P_{-1}(E_+)K_0(E_{+}).
\end{align}
\end{subequations}

The $4\times 4$ recurrences in \Eqs{recurr1} can be reduced to the $2\times
2$ form in the Keldysh space. Indeed, assuming $S_m =
-\sigma_x\overline{S}_m$ and $P_m = -\sigma_x \overline{P}_m$, we arrive at
the recurrences for $\overline{S}$ and $\overline{P}$ which are diagonal in
the Nambu space,
\begin{subequations} \label{recurr2}
\begin{align}
\overline{S}_m = \left(Q_m -i\lambda\sigma_z - q_{m+1}\overline{S}\,'_{m+1}
\right)^{-1}q_m,
\\
\overline{P}_m = ( Q_m -i\lambda\sigma_z -
q_{m}\overline{P}\,'_{m-1})^{-1}q_{m+1},
\end{align}
\end{subequations}
where the prime sign denotes the change of the sign of the
$\sigma_z$-compo\-nent of the matrix. Then the function $K_0$ is also found
to be diagonal in the Nambu space,
\begin{align}
&K_0=\sum_{\sigma=\pm 1} \frac{1}{2}(\sigma+ \sigma_z)\bigl(Q_0
+i\lambda\sigma- q_0 \overline{P}_{-1}^{\;\sigma} - q_1
\overline{S}_1^{\;\sigma} \bigr)^{-1}. \label{K00}
\end{align}
The functions $\overline{S}^{\;\sigma}$ and $\overline{P}^{\;\sigma}$ satisfy
\Eqs{recurr2} in which $\sigma_z$ is replaced by the scalar $\sigma$. Then,
introducing the quantities $s_m(\lambda)$ and $p_m(\lambda)$ according to
$\overline{S}_m^{\;\sigma} = s_m(\sigma\lambda)$ and
$\overline{P}_m^{\;\sigma} = p_m(\sigma\lambda)$, we arrive at the $2\times
2$ recurrences for the Keldysh matrices
\begin{subequations} \label{recurr4}
\begin{align}
s_m = \left[Q_m +i\lambda(-1)^m - q_{m+1}s_{m+1} \right]^{-1}q_m,
\\
p_m = \left[Q_m +i\lambda(-1)^m- q_{m}p_{m-1} \right]^{-1}q_{m+1}.
\end{align}
\end{subequations}
In \Eq{K00} rewritten through the matrices $s_m$ and $p_m$, we can replace
$\sigma\lambda \to \lambda$ that does not change the result of the
integration over $\lambda$ in \Eq{I6}; as the result, only the term
proportional to $\sigma_z$ survives in \Eq{K00}:
\begin{align} \label{K0}
K_0 =\sigma_z K, \quad K=[Q_0 +i\lambda- q_0 p_{-1}(\lambda) - q_1
s_1(\lambda)]^{-1}.
\end{align}
By combining \Eqs{Jm}, \eqref{KK}, and \eqref{K0}, and shifting the energy in
$j_{\pm 1}$ by $\mp eV/2$ which holds the result of integration over $E$ in
\Eq{I6} unchanged, we obtain current spectral densities
\begin{subequations} \label{J01-1}
\begin{align}
j_0 &= 2\Tr\nolimits_\tau K(E,\lambda) G_0^-(E), \label{J0}
\\
j_1 &= -2\Tr\nolimits_\tau p_{-1}(E,\lambda)K(E,\lambda)G_1^-(E_-), \label{Jp1}
\\
j_{-1} &= 2\Tr\nolimits_\tau s_{1}(E,\lambda)K(E,\lambda)G_1^-(E_+).
\label{Jm1}
\end{align}
\end{subequations}

\section{Numerical Results \label{Numerics}}

\begin{figure}[tb]
\centerline{\epsfxsize=7.5cm\epsffile{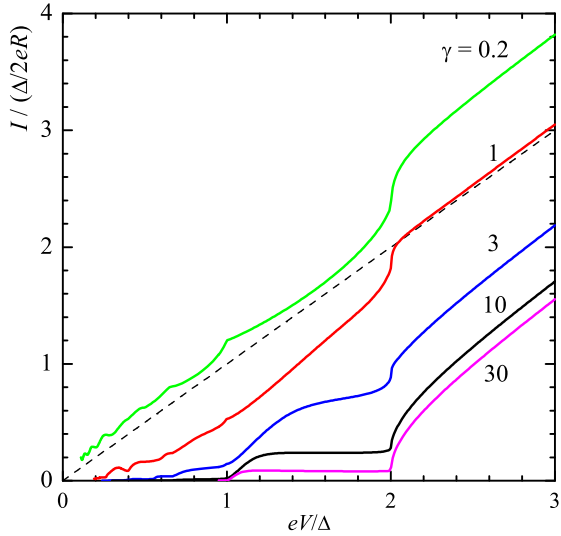}}
\caption{(Color online) Current-voltage characteristics of a short diffusive
$\text{SINIS}$ junction at different values of the parameter $\gamma$.}
\label{sgs}
\end{figure}
\begin{figure}[tb]
\centerline{\epsfxsize=8.0cm\epsffile{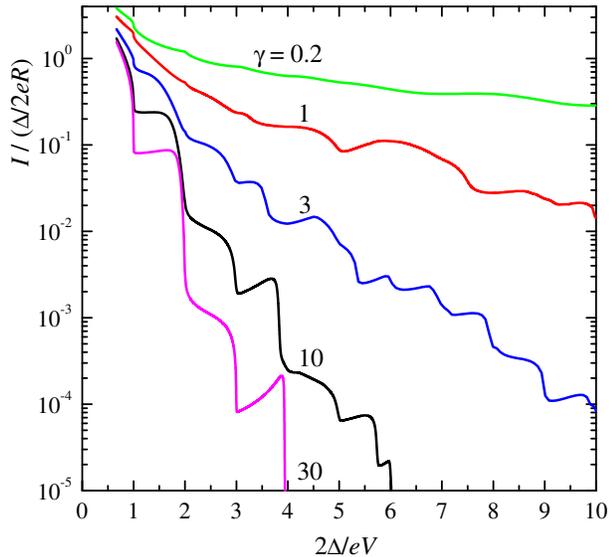}}
\caption{(Color online) Current vs inverse voltage in logarithmic scale.}
\label{sgslog}
\end{figure}
\begin{figure}[tb]
\centerline{\epsfxsize=8cm\epsffile{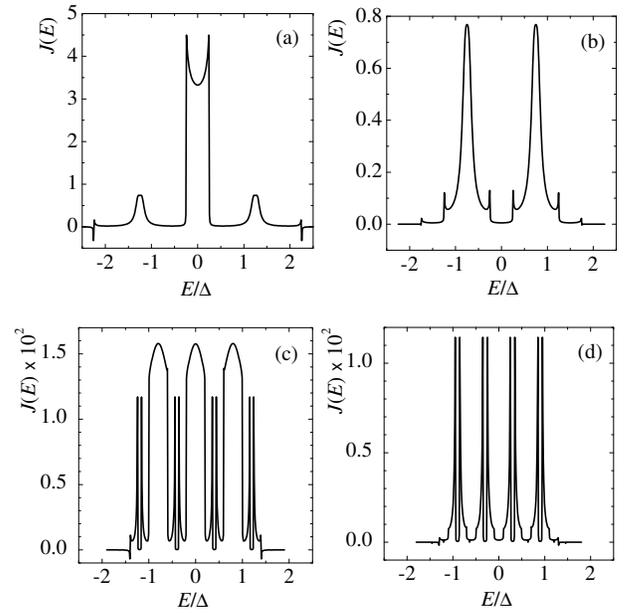}}
\caption{Current spectral density $J(E)$ [normalized on $(2eR)^{-1}$] at
$\gamma = 10$ and different applied voltages: (a) $eV = 2.5\Delta$,
single-particle current; (b) $eV=1.5\Delta$, 2-particle current; (c)
$eV=0.8\Delta$, 3-particle current; (d) $eV=0.6\Delta$, 4-particle current.
Visible are small contributions of higher even-particle processes into the
spectral density of the odd-particle currents. }
\label{currents}
\end{figure}

Now we proceed with the numerical analysis of the dc current using
\Eqs{recurr4}-\eqref{J01-1} and \eqref{I6}.  The results obtained for a wide
range of values of the parameter $\gamma$ and at zero temperature are shown
in Figs.~\ref{sgs} and \ref{sgslog}. As expected, at small electron-hole
dephasing $\gamma = 0.2$, the IVC is close to the results\cite{Naveh,BG}
found from \Eqs{Nazarov} and \eqref{rho}. In this case, the IVC consists of
concave portions between weakly pronounced features (steps and peaks) and
reveals the excess current at large voltages. Since the edges $\pm\Delta_g$
of the minigap $\Delta_g \approx 0.8\Delta$ are close to the edges
$\pm\Delta$ of the superconducting energy gap, the existence of such large
minigap does not distort the periodicity of the IVC features, the positions
of which approximately coincide with the gap subharmonics $eV =2\Delta/n$,
$n=1,2,\ldots$.

As the junction transparency decreases (i.e., $\gamma$ increases), the excess
current also decreases, approaching zero at $\gamma \approx 1$, and then
becomes negative (deficit current). Simultaneously, the peaks in the IVC
almost vanish, and the subharmonic gap structure at small $eV $ and $\gamma =
1 - 3$ becomes somewhat chaotic (see Fig.~\ref{sgslog}). This is due to the
interplay of the contributions of MAR trajectories touching the
superconducting gap edges and the edges of the minigap $\Delta_g \approx
(0.25 - 0.5)\Delta$. Then, at large $\gamma = 10-30$, the IVC features become
regular again and their positions exactly correspond to the gap subharmonics.
In this case, the minigap is small, $\Delta_g \approx 0.03-0.1$, and
therefore affects the MAR trajectories touching the superconducting gap edges
with even number of steps only. The enhanced density of states in the
vicinity of the minigap increases the transmissivity of these MAR chains;
this leads to anomalous enhancement of the magnitude of the dc current just
above the even gap subharmonics $n = 2k$. As $k$ increases, this resonance
effect becomes more pronounced and leads to the appearance of the IVC
portions with negative differential resistance, as seen in Fig.~\ref{sgslog}.
The current spectral density $J(E)$ shown in Fig.~\ref{currents} at
$2\Delta/n < eV < 2\Delta/(n-1)$, $n=1- 4$, has the form of $n$ main equal
peaks which acquire a resonant shape for $n = 2k$ [Fig.~\ref{currents}(b) and
\ref{currents}(d)], in accordance with the above-mentioned anomalous
transmissivity of even MAR chains. Small footprints of these resonances are
visible in the spectral density of the odd-particle currents with $n=2k-1$
[Fig.~\ref{currents}(a) and \ref{currents}(c)].

We note that, at $\gamma > 1$ and $eV < \Delta$, the averaged IVC is well
approximated by the dependence
\begin{equation}
I(V) = \frac{0.18\Delta^2}{eR\Delta_g}\gamma^{-2\Delta/eV},
\end{equation}
which is similar to the result for the ballistic SIS structure
\cite{Bratus97} with $\gamma^{-1}$ standing for the transparency $D$. Thus,
at $\gamma >1$, the quantity $\gamma^{-1}$ plays the role of an effective
parameter for multiparticle tunneling processes which determines the value of
the $n$-particle current of the order of $R^{-1}\gamma^{1-n}$. This
conclusion is confirmed by asymptotic analysis of multiparticle currents (see
below).

\section{Analytical results \label{Analytics}}

In this section, we present a detailed analytical discussion of the dc
current. To this end, we consider \Eqs{recurr4} as functional equations for
the functions $S(E,\lambda)=  s_{ 1}G^+_1(E_+)$ and $P(E,\lambda) =
p_{-1}G^+_1(E_-)$,
\begin{align}\label{PS}
&\!\!\!\!(S,P)(E,\lambda)\! =\left[K ^0(E\pm eV,-\lambda) +\Pi^{S,P}(E\pm
eV,-\lambda)\right]^{-1}\!\!\!\!\!,
\\
&K^0(E,\lambda)=G^+_0-i\tau_d E+i\lambda,
\\ \label{Pi}
&\Pi^{S,P}(E,\lambda)=-G^+_1(E_\pm)(S,P)(E,\lambda)G^+_1(E_\pm).
\end{align}
In terms of these functions, the contributions \Eq{Jp1} and \eqref{Jm1} of
nonzero harmonics to the dc current read as
\begin{align}
&j_{\pm 1} =\mp 2 \Tr_\tau (P,S) G^+_1(E_\mp)\widetilde{K}^{-1} G^-_1(E_\mp),
\\
&\widetilde{K}= K ^0+\Pi^P +\Pi^S.\label{tK}
\end{align}

Analysis of \Eq{PS} shows that the matrices $K^0$, $P$, and $S$ possess
certain symmetry pro\-perties with respect to the transformation $E \to -E$,
$\lambda \to -\lambda$: $P^{1,+}\leftrightarrow -S^{1,+}$,
$P^z\leftrightarrow S^z$, $Z_P\leftrightarrow Z_S$, $Z_K\leftrightarrow Z_K$,
where $Z_K = \det\widetilde{K}$, $Z_P =\det P$, $Z_S = \det S$, and the upper
indices $1$, $z$, and $+$ denote $1$-, $\tau_z$-, and $\tau_+$-matrix
components, respectively. These relations allow us to exclude the
antisymmetric terms that vanish under integration over $E$ and $\lambda$ in
\Eq{I6} and to write down the spectral densities \Eqs{J01-1} in a compact
form
\begin{subequations} \label{J2}
\begin{align}
j_0 &=-\frac{2}{Z_K}\bigl[
2\widetilde{K}^z(N_+f_+-N_-f_-)-\widetilde{K}^+(N_+ -N_- )\bigr],\label{J-0}
\\
j_1 &+j_{ -1} =-\frac{2}{Z_K}   \overline{M}^2 _-  \bigl(P^z
\widetilde{K}^+-P^+\widetilde{ K}^z \bigr). \label{J-1-1}
\end{align}
\end{subequations}
\subsection{Excess current at $eV\gg \Delta$}

We start with evaluation of the excess current $I^{exc}$ at large applied
voltage and for arbitrary $\gamma$. This quantity is contributed by both the
single-particle current and the two-par\-ticle Andreev current. Formally,
$I^{exc}$ is the voltage-in\-de\-pen\-dent term in asymptotic expression for
the dc current $I = V/2R + I^{exc}+O(\Delta/eV)$ at $eV \gg \Delta$. In
corresponding expansion of full current spectral density \Eq{J2}, we truncate
the recurrences for $P$ and $S$, omitting in $j(E,\lambda)$ the combinations
of the functions $M_\alpha$ or $\oM_\alpha$ with different energies (such as
$M_\alpha M_\beta$ with $\beta \neq \alpha$), which turn to zero at $eV \to
\infty$,
\begin{align}\label{jexc}
&j(E,\lambda) =   \frac{Z_P}{Z_K} \biggl\{ 2\oM^2_{-1/2} L_{1/2,-3/2}- M
^2_{-1/2}\biggl[ L_{1/2,-1/2}
\\
&\times\biggl(1+\frac{N_{-3/2}}{N_{-1/2}}\biggr) +
L_{-1/2,-3/2}\biggl(1-\frac{N_{1/2}}{N_{-1/2}}\biggr) \biggr]\biggr\}-4\frac{
L_{1/2,-1/2}}{Z_K}, \nonumber
\\
&L_{\alpha \beta}=N_\alpha N_\beta(f_\alpha - f_\beta),\quad \alpha >
\beta.\label{L}
\end{align}
Here and in the following, the lower indices denote the energy shift, e.g.,
$N_\alpha \equiv N(E+\alpha eV)$. At zero temperature ($f = \sgn E$), the
factor $L_{\alpha\beta}$ is nonzero within the energy region
\begin{equation}\label{domain}
\Delta - \alpha eV < E  < -\Delta -\beta eV .
\end{equation}
Existence of this energy interval, in which $L_{\alpha\beta}= 2N_\alpha
N_\beta$, imposes the following condition on the applied voltage,
\begin{equation}\label{eV}
(\alpha - \beta) eV > 2\Delta;
\end{equation}
otherwise, the function $L_{\alpha\beta}$ is identically zero. Since the
convergence of integration over $E$ of all terms in curly brackets in
\Eq{jexc} is ensured by the functions $M_\alpha$ or $\oM_\alpha$ with
$E_\alpha \sim \Delta$, the functions $N_\beta$ with different energies
($\beta \neq \alpha$) can be approximated by their values in a normal metal
$N_\beta = 1$. By using these simplifications, we arrive at the following
asymptotics of the dc current,
\begin{align}\label{Iinfty.1}
&I = \frac{2\Delta}{\pi eR} \int^\infty_{-\infty} \!\!\!\!d\lambda
\left[\int^{eV/2\Delta}_{1}dx \,j_1(x,\lambda) + \int^{ 1}_{0}dx
\,j_2(x,\lambda) \right],
\\
&j_1=\frac{x[(\lambda +y )^2+ z^2-1]} {|\lambda^2-1+(z -iy)^2|^2}, \quad j_2=
\frac{ 2y_1}{ | \lambda^2 +1 - (z_1 -iy_1)^2|^2},\nonumber
\\
&y=2\gamma x\sqrt{x^2-1}, \quad z = x+\sqrt{x^2-1},\nonumber
\\
&y_1=\sqrt{1-x^2}, \quad z_1=x(1+2\gamma y_1). \nonumber
\end{align}
By integrating over $\lambda$ in \Eq{Iinfty.1} and separating out the
constant term, we obtain a general expression for the excess current,
\begin{align} \label{excess}
&I^{\textit{exc}}(\gamma) = \frac{\Delta}{eR} \left[
\int^{\infty}_{1}dx\,j_1(x) + \int^{ 1}_{0}dx\, j_2(x) -1\right],
\\
&j_1(x)=\frac{\sqrt{2} x
(\widetilde{T}+T^2-1)}{\widetilde{T}\sqrt{\widetilde{T}-A} }-1, \quad
j_2(x)=\frac{ 2 \sqrt{2(1-x^2)}}{{T_1}\sqrt{{T}_1+A_1} }, \nonumber
\\
&T^2={y^2+z^2},\quad \widetilde{T}^2={A^2+4y^2z^2}, \quad A=1+y^2-z^2,
\nonumber
\\
&{T}_1^2= {A_1^2+4y_1^2z_1^2},\quad A_1= 1+y_1^2-z_1^2.\nonumber
\end{align}

In the limit of small dephasing $\gamma \to 0$, when the integral over $x$ in
\Eq{Iinfty.1} can be explicitly calculated, the substitution
$D=(\lambda^2+1)^{-1}$ leads to the formula\cite{BG}
\begin{align} \label{Iexc.gamma0}
&I^{\textit{exc}} = \frac{\pi}{2eR} \int^1_{0} {dD }\;\rho(D)
I^{\textit{exc}}_{\textit{SIS}}(D) = 0.53\frac{\Delta}{eR}, \quad \gamma \ll
1,
\\
&I^{\textit{exc}}_{\textit{SIS}}(D)=\frac{D^2\Delta}{\pi\mathcal{R}} \left[
1- \frac{D^2}{2(1+\mathcal{R})\sqrt{\mathcal{R}}}
\ln\frac{1+\sqrt{\mathcal{R}} }{1-\sqrt{\mathcal{R}} }\right], \;\;
{\mathcal{R}}=1-D,\nonumber
\end{align}
which expresses the excess current through its value
$I^{\textit{exc}}_{\textit{SIS}}$ for a single ballistic channel\cite{LTP97}
averaged over the transparency distribution \Eq{rho}, in accordance with
\Eq{Nazarov}. In the opposite case $\gamma \gg 1$, $I^{\textit{exc}}$ becomes
negative (deficit current),
\begin{eqnarray}
I^{\textit{def}} = -\frac{2\Delta}{3eR}, \qquad \gamma \gg 1.
\end{eqnarray}

\subsection{Multiparticle currents at large $\gamma$}

In the limit of large dephasing $\gamma \gg 1$, it is possible to express
analytically the full current as a sum of contributions of $n$-particle
tunneling processes.\cite{MAR95-1} Here, we proceed with the asymptotic
analysis of these partial contributions. First, we separate out the unity and
the traceless components of the Keldysh matrices, e.g., $P=P^1+\hat{P}$,
$\hat{P}\equiv \tau_z P^z+\tau_+P^+$,
\begin{subequations} \label{defPi}
\begin{align}
\hat{\Pi}^{S,P} &=-G_{\pm 0}(\hat{S},\hat{P}) G_{\pm 0}, \quad \Pi^{S\, 1,P\,
1}= \tM_{\pm 1/2}(S^1,P^1),
\\
\tM_{\alpha} &= -\left[G^+_1(E_\alpha)\right]^2
=\frac{1}{4}\bigl(\oM^2_{\alpha}-M^2_{\alpha}\bigr),\label{tM}
\end{align}
\end{subequations}
where $G_{\pm m} = G^+_1(E_\pm \pm meV)$. By introducing the notations
\begin{align}
\hat{K}^0_{m} &\equiv\hat {K^0}(E_{m} ),\quad K^{01}_{m}\equiv K^{01}[ E_{m},
(-1)^m\lambda ],\nonumber
\\
Z_{\pm m} &\equiv \det\left\{\hat{K}^0_{\pm m}+K^{01}_{\pm m}+ \Pi^{S,P}
[E_{\pm m},(-1)^m\lambda ]\right\}, \quad m>0,\nonumber
\end{align}
we rewrite \Eqs{PS} in an expanded form, explicitly performing the
recurrences for the functions $P$ and $S$. The result can be presented in the
form of the series for the functions $\Pi$,
\begin{subequations} \label{SeriesPi}
\begin{align}
&\hat{\Pi}^S = \sum_{m=1}^\infty \hat{\Pi}^S_m, \quad \Pi^{S1} =
\sum_{m=1}^\infty K^{01}_{m}\prod_{\alpha=1}^{m}\tM_{\alpha-1/2}
Z^{-1}_{\alpha}, \label{PiP}
\\
&\hat{\Pi}^S_m = G_{+0}G_{1}...G_{m-1}\hat{K}^0_{m}
G_{m-1}...G_{1}G_{+0}\prod_{\alpha=1}^{m} Z^{-1}_{\alpha}.\label{PiPn}
\end{align}
\end{subequations}
The series for ${\Pi}^P$ differ from \Eqs{SeriesPi} by opposite signs of all
lower indices. With this remark, the series for the functions $P$ and $S$ can
be obtained from \Eqs{SeriesPi} and \eqref{defPi}. These series can be
interpreted as asymptotic expansions over $\gamma^{-1}$, due to the presence
of large parameter $\gamma \gg 1$ in $Z_\alpha$. Physically, these expansions
reflect the nature of the net current as a sum of $n$-par\-ticle tunnel
currents;\cite{MAR95-1} each of them exists at $eV > 2\Delta/n$ and scales as
$\gamma^{1-n}$ with respect to the single-par\-ticle current. The latter fact
allows us to consider the $n$-par\-ticle current only within its actual
voltage region $2\Delta/n < eV < 2\Delta/(n-1)$; at larger voltages, the
$(n-1)$-par\-ticle current dominates. Estimation shows that $m$th terms in
\Eqs{SeriesPi} contribute to the $(m+1)$-particle current; thus, it is enough
to consider them only at $eV < 2\Delta/m$, which greatly simplifies the
structure of the series.

Indeed, consider the term $\hat{\Pi}^S_1$ proportional to the product
\begin{align} \label{GKG}
G_{+0}\hat{K}^0_{1}G_{+0} = -\tM_+ \hat{K}^0_{1} -\frac{1}{2} N_\alpha
M^2_\beta  (f_\alpha - f_\beta)\tau_+,
\end{align}
where $\alpha = 3/2$, $\beta = 1/2$, and we used the identities $F_\alpha^2 =
1$ and $F_\alpha F_\beta F_\alpha = \tau_z +(4f_\alpha - 2f_\beta)\tau_+$.
The allowed energy region determined by the last term in \Eq{GKG} is similar
to that for the function $L_{\alpha\beta}$ [see \Eqs{L} and \eqref{domain}],
therefore this term turns to zero at $eV < 2\Delta$, according to \Eq{eV}.
Thus, in this voltage region, the action of matrix envelopes $G_{+0}$ on the
matrix $\hat{K}^0_{1}$ is reduced to multiplication on the scalar factor
$-\tM_+$. Similar considerations applied to each term of the expansion
\Eq{PiP} lead to the following simplified series for the functions
$\hat{\Pi}$,
\begin{align}
\label{PiP1} \hat{\Pi}^S= \sum_{m=1}^\infty \theta (2\Delta - m eV) (-1)^m
\hat{K}^0 _{m} \prod_{\alpha =1}^{m}{\widetilde{M}_{\alpha,
-1/2}}{Z_{\alpha}^{-1}},
\end{align}
where we introduced the Heaviside step function $\theta$ to specify
explicitly the relevant voltage regions. The series for the functions
$\hat{\Pi}^P$ differ from \Eq{PiP1} by opposite signs of lower indices.

Now we proceed with asymptotic evaluation of the dc current. First we
consider the contribution $j_0$ in \Eq{J-0} to the net current spectral
density which, according to \Eq{tK}, can be presented as a sum of three
terms,
\begin{align}\label{J02} &j_0 = j_0^{K}+j_0^{P}+j_0^{S},
\\
&j_0^{K} =-\frac{2}{Z_K}\left[ 2 K _0^z(N_+f_+-N_-f_-)-K _0 ^+(N_+ -N_-
)\right]. \label{J0main}
\end{align}
The first term is equal to $-(4/Z_K)L_{1/2,-1/2}$ and represents the spectral
density of the sin\-gle-par\-ticle current. According to \Eqs{domain} and
\eqref{eV}, it exists within the energy interval $|E|<-\Delta +eV/2$ and
vanishes at $eV < 2\Delta$. Thus, at subgap voltages, we have to involve the
terms $j_0^{P,S}$, which differ from \Eq{J0main} by replacements
$K_0\to\Pi^{P,S}$. Considering, e.g., the spectral density $j_0^S$ and taking
into account \Eq{PiP1}, we found that the $m$th term in the expansion of
$j_0^S$ is proportional to
\begin{align}
&\theta(2\Delta-meV)\sum_{\alpha,\beta = \pm 1/2}\beta L_{m+\alpha,
\beta}.\label{LLLL}
\end{align}
As follows from \Eq{eV}, only the term with $\alpha= -\beta=1/2$,
proportional to $\theta[(m+1)eV-2\Delta]$, survives in \Eq{LLLL}. Thus, we
obtain the following series for $j_0^S$,
\begin{align}\label{J0Pexp1}
&\!\!\!\!j_0^{S}=\frac{2}{Z_K}\sum_{m=2}^\infty \chi_{m}(V) L_{m-1/2,-1/2}
\prod_{\alpha=1}^{m-1} \frac{a_{\alpha-1/2}}{|Z_{\alpha}|}, \;\; a_\alpha
=\frac{\oM^2_\alpha}{4},
\\
&\chi_m(V) = \begin{cases} 1, & 2\Delta/m < eV < 2\Delta/(m-1),
\\ 0 & \text{otherwise}. \end{cases}  \nonumber
\end{align}
By applying similar considerations to $j_0^P$ and $j_1+j_{-1}$, we arrive at
the formula for full generalized current spectral density [at $n=1$, the
product in \Eq{Jnfin} is unity],
\begin{align}
&j(E,\lambda)=\sum_{n=1}^\infty \chi_n(V) j^{(n)},
\\
&j^{(n)}=\frac{4 }{|Z_K|} \sum_{m=1}^n L_{m-1/2,m-n-1/2}
\prod_{\alpha=m-n+1}^{m-1}\frac{a_{\alpha-1/2}}{|Z_{\alpha}|}. \label{Jnfin}
\end{align}

According to \Eq{Jnfin}, the $n$-particle spectral density $j^{(n)}$ consists
of $n$ equal contributions of MAR chains with $n$ steps. Each chain starts at
the energy $E_{m-n-1/2} < -\Delta$ and finishes at $E_{m-1/2}>\Delta$, thus
transferring the quasiparticles to the extended states above the energy gap,
which results in formation of the dissipative current. The intermediate
energies $E_{\alpha-1/2}$ inside the gap correspond to the points of Andreev
reflections. These contributions are nonzero within the energy intervals of
width $neV - 2\Delta$, which are distributed equidistantly (with spacing
$eV$) along the energy axis and symmetrically with respect to the zero
energy, in conformity with the numerical results shown in
Fig.~\ref{currents}. This enables us to write down the full dc current as a
sum of $n$-particle tunnel currents $I^{(n)}$, where only one term in
$j^{(n)}$ multiplied by $n$ is taken into account,
\begin{align}\label{fullcurrent1}
&I=\sum_{n=1}^\infty \chi_n(V) I^{(n)},
\\
&I^{(n)}={n} \int^\infty_{-\infty}\frac{d\lambda}{2\pi} \int^{-\Delta
+(n-1/2)eV }_{\Delta-eV/2} \frac{dE}{eR} \,\frac{N_{1/2} N_{1/2-n}} {|Z_K|}
\prod_{\alpha =1}^{n-1}\frac{a_{1/2-\alpha}} {|Z_{-\alpha}|}, \label{In}
\end{align}
At $n=1$, the product in Eq.\;\eqref{In} is assumed to be unity.

To complete our consideration, we present final expressions of $n$-particle
currents for $n=1,2$, and $3$ obtained from \Eq{In} by the integration over
$\lambda$. A nontrivial point in this procedure is a proper choice of
approximation for the determinants $Z$. In the single-particle current, it is
enough to take $Z_K$ in the main approximation as $\det K^0=-[(\lambda
-\tau_d E)^2+(1/4)(N_++N_-)^2]$, neglecting contributions of $\Pi^{P,S}$ to
the function $\widetilde{K}$ in \Eq{tK}. As is obvious from this expression,
the parameter $\tau_d$ drops out from $I^{(1)}$, and we obtain a simple
formula
\begin{equation}\label{I1-1}
I^{(1)}= \int^{-\Delta +\frac{eV}{2}}_{\Delta -\frac{eV}{2}} \frac{
dE}{eR\bigl(N_+^{-1} + N_-^{-1}\bigr)}.
\end{equation}
From the standpoint of the circuit theory for incoherent $\text{SINIS}$
structures,\cite{Bezugly2000} this result can be interpreted as the Ohm's law
for two tunnel resistors $R_\pm = R N^{-1}_\pm$ connected in series.

In calculation of the two-particle current, the main approximation is
applicable to the determinant $Z_{-1}=-|X_{-1}|^2$, whereas in
$Z_K=-|X_{K}|^2$ one should hold the term $\Pi^{P}$,
\begin{align}
& X_{-1}=i\lambda_{-1}+\frac{1}{2}N_{-3/2}, \quad X_K =i\lambda_{K}
+\frac{1}{2}N_{+}+\frac{a_-}{X_{-1}},\nonumber
\\
&\lambda_K = \lambda -\tau_d E +\frac{1}{2}\oN_{-}, \quad \lambda_{-1} =
-\lambda_K-2\tau_d E_-+\oN_{-}.\nonumber
\end{align}
By taking $\lambda_K$ as a new integration variable, we see that its
characteristic value is of the order of unity, which enables us to
approximate $\lambda_{-1}$ as $-2\tau_d E_-$. After integration over
$\lambda_K$ and symmetrization of the allowed energy interval, we obtain
\begin{align}
&I^{(2)}= \int^{-\Delta + eV}_{\Delta - eV } \frac{dE}{eR} \frac{8N_1 a
N_{-1}}{ N_1  [ (4\tau_d E)^2 + N_{-1}^2]+4a N_{-1}}. \label{I2-1}
\end{align}
The current spectral density in \Eq{I2-1} has a resonant form, with a sharp
peak at zero energy (the resonant nature of the even-par\-ticle currents has
been already noted in Sec.~\ref{Numerics}). If the applied voltage is not
very close to the threshold $\Delta/e$ of the two-par\-ticle current, the
integral in \Eq{I2-1} can be calculated in the resonant approximation by
assuming $E=0$ in all spectral functions and spreading the limits to $\pm
\infty$,
\begin{equation}\label{I2-2}
I^{(2)}=\frac{\pi\Delta}{2\gamma eR} \frac{ N(eV)}{\sqrt{1+N^2(eV)}}.
\end{equation}
Similar considerations lead to the following expression for the
three-particle current,
\begin{align}\label{I3}
I^{(3)} =\frac{3}{4\gamma^2 eR}\int^{-\Delta +\frac{3}{2}eV}_{\Delta
-\frac{3}{2}eV} \frac{dE\; N_{ \frac{3}{2}} a_{ +} a_{- }
N_{-\frac{3}{2}}}{a_{- }N_{-\frac{3}{2}}E^2_+ +a_{+}N_{ \frac{3}{2}}E^2_- }.
\end{align}
Expressions \Eqs{I1-1}--\eqref{I3} well reproduce the results of full
numerical calculations for large $\gamma$ shown in Figs.~\ref{sgs} and
\ref{sgslog}. \vspace{2mm}

\section{Summary}

We have calculated, both numerically and analytically, the cur\-rent-voltage
characteristics (IVCs) of a diffusive $\text{SINIS}$ junction, where S are
local-equilibrium superconducting re\-ser\-voirs, I denotes tunnel barriers,
and N is a short diffusive normal wire, the length of which is much smaller
than the coherence length and the resistance $R_N$ is much smaller than the
resistance $R$ of the tunnel barriers. The regime of coherent MAR transport
in such structure is governed by the parameter $\gamma = \tau_d \Delta$,
which represents a characteristic phase shift between the wavefunctions of
the electron and the retro-reflected hole accumulated during the
quasiparticle dwell time $\tau_d \sim E_{\text{Th}}^{-1} (R/R_N)$. We
demonstrated that the Keldysh-Green's function equations for this problem can
be efficiently solved in the whole range of electron-hole dephasing
$0<\gamma<\infty$. This is achieved by reducing the solution of full $4\times
4$ matrix two-time Keldysh-Green's function equations\cite{LO} to the
solution of the $2\times 2$ matrix recurrence relations of the second order,
similar to the recurrences in analogous ballistic
problems.\cite{MAR95-1,MAR95-2} In the limit of small dephasing $\gamma \to
0$, our solution reduces to a known formula for mesoscopic
connector,\cite{Nazarov99} i.e., averaging of the result for the single
channel junction over the distribution of transparencies for the
corresponding double-barrier normal diffusive structure.\cite{BG}

In the opposite case of large electron-hole dephasing $\gamma \gg 1$, the
subharmonic gap structure in the IVC scales with $\gamma^{-1}$; this means
that the $n$-particle tunnel currents scale as $\gamma^{1-n}$ with respect to
the single-particle current, and $\gamma^{-1}$ plays the role of an effective
tunneling parameter. Due to the presence of resonant MAR chains touching the
edges of small minigap $\Delta_g \approx\Delta/(\gamma+1)$, the even
subharmonics are enhanced, and corresponding portions of the IVC show
negative differential resistance. We presented analytical results for the
excess current at arbitrary $\gamma$ and for multiparticle currents at
$\gamma \gg 1$.

For experimental observation of the phenomena discussed in this paper, the
most stringent constraint concerns the Jo\-seph\-son regime. This implies
small values of the dwell time compared to the inelastic relaxation time. If
this requirement is not fulfilled the central metallic island acts as a
reservoir, and the structure splits in two NIS junctions connected in series.
This is the case of SINIS junctions extensively used in microcoolers
\cite{Giazotto} and SET turnstiles.\cite{Pekola} For the Josephson effect to
occur in metallic SINIS junctions with conventional oxide tunnel barriers, a
sandwich-type junctions must be employed having extremely thin normal
metallic layer not exceeding 10 nm. Such junctions have been developed using
Nb/AlO$_x$/Al/AlO$_x$/Nb technology, and they demonstrated rather large
values of $\gamma\sim 10^4$ and a pronounced deficit
current.\cite{mixer,Ketter,Twente} This is precisely the limit of large
electron-hole dephasing studied in this paper. In order to investigate a
crossover to the regime of small dephasing at $\gamma\sim 1$ one needs to use
junctions with more transparent NS interfaces, such as junctions based on
diffusive InAs 2D electron gas or graphene, or corresponding nanowires and
nanotubes.


\begin{thebibliography}{99}

\bibitem{KBT}
T.~M.~Klapwijk, G.~E.~Blonder, and M.~Tinkham, Physica B \& C {\bf 109-110},
1657 (1982).

\bibitem{Arnold}
G.~B.~Arnold, J. Low Temp. Phys. {\bf68}, 1 (1987).

\bibitem{Zaikin}
U.~Gunsenheimer and A.~D.~Zaikin, Phys. Rev. B {\bf 50}, 6317 (1994).

\bibitem{MAR95-1}
E.~N.~Bratus', V.~S.~Shumeiko, and G.~Wendin, Phys.\ Rev.\ Lett. {\bf 74},
2110 (1995).

\bibitem{MAR95-2}
D.~Ave\-rin and A.~Bardas, Phys.\ Rev.\ Lett. {\bf 75}, 1831 (1995).

\bibitem{Cuevas96}
J.~C.~Cuevas, A.~Mart\'in-Rodero, and A.~Levy Yeyati, Phys. Rev. B {\bf 54},
7366 (1996).

\bibitem{Bardas}
A.~Bardas and D.~V.~Averin, Phys. Rev. B {\bf 56}, R8518 (1997).

\bibitem{Naveh}
Y.~Naveh, Vijay Patel, D.~V.~Averin, K.~K.~Likharev, and J.~E.~Lu\-kens, {
Phys. Rev. Lett.} {\bf 85}, 5404 (2000).

\bibitem{BG}
A.~Brinkman and A.~A.~Golubov, {Phys.\ Rev.\ B} {\bf 61}, 11297 (2000).

\bibitem{Ingermann2001}
\AA.~Ingerman, G.~Johansson, V.~S.~Shumeiko, and  G.~Wendin, Phys. Rev.  B
{\bf 64}, 144504 (2001).

\bibitem{Cuevas97}
A.~Levy Yeyati, J.~C.~Cuevas, A.~L\'opez-D\'avalos, and A.~Mart\'in-Rodero,
Phys. Rev. B {\bf 55}, R6137 (1997).

\bibitem{Johansson99}
G.~Johansson, E.~N.~Bratus', V.~S.~Shumeiko, and G.~Wendin, Phys. Rev. B {\bf
60}, 1382 (1999).

\bibitem{Samuelsson2002}
P.~Samuelsson, G.~Johansson, \AA.~Ingerman, V.~S.~Shumeiko, and  G.~Wendin,
Phys. Rev.  B {\bf 65}, 180514(R) (2002).

\bibitem{LO} A.~I.~Larkin and Yu.~N.~Ovchinnikov, in {\em
Nonequilibrium Superconductivity}, edited by D.~N.~Lan\-gen\-berg and
A.~I.~Lar\-kin (Elsevier, Amsterdam, 1986).

\bibitem{Zaitsev}
A.~V.~Zaitsev and D.~V.~Averin, Phys. Rev. Lett. {\bf 80}, 3602 (1998).

\bibitem{Bezugly2000}
E.~V.~Bezuglyi, E.~N.~Bratus', V.~S.~Shumeiko, G.~Wendin, and
H.~Ta\-kayanagi, Phys.\ Rev.\ B {\bf 62}, 14439 (2000).

\bibitem{Bezuglyi2001}
E.~V.~Bezuglyi, E.~N.~Bratus', V.~S.~Shumeiko, and G.~Wendin, Phys. Rev. B
{\bf 63}, 100501(R) (2001).

\bibitem{OTBK}
M.~Octavio, M.~Tinkham, G.~E.~Blonder, and T.~M.~Klapwijk, Phys. Rev. B {\bf
27}, 6739 (1983).

\bibitem{Bezuglyi2006}
E.~V.~Bezuglyi, A.~S.~Vasenko, E.~N.~Bratus', V.~S.~Shumeiko, and G.~Wendin,
Phys. Rev. B, {\bf 73}, 220506 (2006).

\bibitem{Cuevas06}
J.~C.~Cuevas, J.~Hammer, J.~Kopu, J.~K.~Viljas, and M.~Eschrig, Phys.\ Rev.\
B {\bf 73}, 184505 (2006).

\bibitem{Volkov}
R.~Seviour and A.~F.~Volkov, Phys. Rev. B {\bf 61}, 9273 (2000).

\bibitem{Brinkman}
A.~Brinkman, A.~A.~Golubov, H.~Rogalla, F.~K.~Wilhelm, and M. Yu.~Kupriyanov,
{Phys.\ Rev.\ B} {\bf 68}, 224513 (2003).

\bibitem{3T}
E.~V.~Bezuglyi, V.~S.~Shumeiko, and G.~Wendin, Phys.\ Rev.\ B {\bf 68},
134506 (2003).

\bibitem{Nazarov99}
Yu.~V.~Nazarov, {Superlatt.\ Microstruct.} {\bf 25}, 1221 (1999); W.~Bel\-zig
and Yu.~V.~Naza\-rov, {Phys.\ Rev.\ Lett.} {\bf 87}, 197006 (2001).

\bibitem{Naz2B}
J.~A.~Melsen and C.~W.~J.~Beenakker, Physica B {\bf 203}, 219 (1994);
W.~Bel\-zig, A.~Bra\-taas, Yu.~V.~Nazarov, and G.~E.~W.~Bauer, { Phys.\ Rev.\
B} {\bf 62}, 9726 (2000).

\bibitem{KL}
M.~Yu.~Kupriyanov and V.~F.~Lukichev, Zh. Eksp. Teor. Fiz. {\bf 94}, 139
(1988) [Sov.\ Phys.\ JETP {\bf 67}, 1163 (1988)].

\bibitem{Samuel}
P.~Samuelsson, {Phys. Rev. B} {\bf 67}, 054508 (2003).

\bibitem{Solution}
J.~B\"{o}rlin, W.~Belzig, and C.~Bruder, {Phys. Rev. Lett.} {\bf 88}, 197001
(2002).

\bibitem{Schep}
K.~M.~Schep and G.~E.~W.~Bauer, {Phys. Rev. Lett.} {\bf 78}, 3015 (1997);
{Phys. Rev. B} {\bf 56}, 15860 (1997).

\bibitem{Bratus97}
E.~N.~Bratus', V.~S.~Shumeiko, E.~V.~Bezuglyi, and G.~Wendin, {Phys. Rev. B}
{\bf 55}, 12666 (1997).

\bibitem{LTP97}
V.~S.~Shumeiko, E.~N.~Bratus', and G.~Wendin, Low Temp. Phys. {\bf 23}, 181
(1997).

\bibitem{Giazotto}
F.~Giazotto, T.~T.~Heikkil\"a, A.~Luukanen, A.~M.~Savin, and J.~P. Pekola,
Rev. Mod. Phys. {\bf 78}, 217 (2006).

\bibitem{Pekola}
J.~P.~Pekola, V.~F.~Maisi, S.~Kafanov, N.~Chekurov, A.~Kemppinen,
Yu.~A.~Pashkin, O.-P.~Saira, M.~M\"ott\"onen, and J.~S.~Tsai, Phys. Rev.
Lett. {\bf 105}, 026803 (2010).

\bibitem{Twente}
E.~Bartolom\'e, A.~Brinkman, J.~Flokstra, A.~A.~Golubov, H.~Rogalla, Physica
C {\bf 340}, 93 (2000).

\bibitem{mixer}
M.~M.~Th.~M.~Dierichs, P.~Dieleman, J.~J.~Wezelman, C.~E.~Honingh, and
T.~M.~Klapwijk, Appl. Phys. Lett. {\bf 64}, 921 (1994).

\bibitem{Ketter}
I.~P.~Nevirkovets, J.~B.~Ketterson, S.~Lomatch, Appl. Phys. Lett. {\bf 74},
1624 (1999).

\end{thebibliography}
\end{document}